\documentclass[aps,prb,twocolumn,letterpaper]{revtex4}
\usepackage{graphicx}
\usepackage{dcolumn}
\usepackage{amsmath}
\usepackage{amssymb}
\usepackage[caption=false]{subfig}
\usepackage{soul}
\usepackage{hyperref}

\newcommand{\subfigimg}[3][,]{%
  \setbox1=\hbox{\includegraphics[#1]{#3}}
  \leavevmode\rlap{\usebox1}
  \rlap{\hspace*{0pt}\raisebox{\dimexpr\ht1-2\baselineskip}{#2}}
  \phantom{\usebox1}
}

\usepackage[usenames,dvipsnames]{color}
\definecolor{orange}{rgb}{1,0.5,0}
\usepackage{float}
\usepackage{epstopdf}
\usepackage{ulem}

\def\kv{{\bf k}}

\def\beq{\begin{equation}}
\def\eeq{\end{equation}}

\def\beqa{\begin{eqnarray}}
\def\eeqa{\end{eqnarray}}

\begin{document}

\title{Anderson Topological Superconductor}
\author{Jan Borchmann, Aaron Farrell and T. Pereg-Barnea}
\affiliation{Department of Physics and the Centre for Physics of Materials, McGill University, Montreal, Quebec,
Canada H3A 2T8}
\date{\today}
\begin{abstract}
In this paper we study the phase diagram of a disordered, spin-orbit coupled superconductor with $s$-wave or $d+id$-wave pairing symmetry in symmetry class $D$. We analyze the topological phase transitions by applying three different methods which include a disorder averaged entanglement entropy, a disorder averaged real-space Chern number, and an evaluation of the momentum space Chern number in a disorder averaged effective model.  We find evidence for a disorder-induced topological state.  While in the clean limit there is a single phase transition from a trivial phase with a Chern number $C=4$ to a topological phase with $C=1$, in the disordered system there is an intermediate phase with $C=3$.  The phase transition from the trivial $C=4$ phase into the intermediate phase with $C=3$ is seen in the real-space calculation of the Chern number. In spite of this, this phase transition is not detectable in the entanglement entropy.  A second phase transition from the disorder induced $C=3$ into the $C=1$ phase is seen in all three quantities. 
\end{abstract}
\maketitle

\section{Introduction} 
Symmetry protected topological\cite{gu,pollmann} (SPT) systems include the quantum spin Hall state, topological insulators in two and three dimensions as well as topological superconductors. These systems, which are generally described by models with multiple phases, share the property that they experience distinct phases which cannot be smoothly transformed into each other while preserving a certain symmetry. In the topological phases, unique properties such as anomalous magneto-resistance and edge/surface states are the result of the topology.  This topology is characterized by topological invariants which are the discreet expectation values of non-local operators. When parameters change across a phase transition the bulk gap closes, allowing the topological invariants to change their values. In particular, in a clean, non-interacting lattice system one can define the Berry curvature in momentum space and integrate it over a relevant area, such as the Brillouin zone. This integral yields the Chern number in broken time reversal symmetry states or a $\mathbb{Z}_2$-invariant in time reversal symmetric states.

While surface states are protected against weak perturbation by the topology, a strongly disordered system can be classified differently than its clean counterpart.  It is therefore interesting to study SPT systems in the presence of disorder.   An example for such a change in classification can be found in two dimensional\cite{groth,xie,Li} and three dimensional\cite{Guo} Anderson topological insulators. In these systems the disorder can be thought of as renormalizing the parameters of the clean system and thus driving the system across topological phase boundaries.  Moreover, while in the clean system the gap in the spectrum is crucial for preventing surface states from scattering into the bulk, in a disordered system, it is the mobility gap which plays this role.

From the point of view of identifying a topological phase transition, disorder poses a challenge.  The introduction of disorder breaks translation invariance and consequently the usual method of computing a topological invariant is invalid as it relies on the existence of a Brillouin zone.  
Alternative approaches, which do not rely on translation invariance, involve integrals over twisted boundary conditions\cite{niu,yang}. These integrals involve a large number of real-space Hamiltonian diagonalizations and consequently are very numerically costly. Efficient alternatives use the same principle and define the Chern numbers via traces\cite{kitaev06} or commutators\cite{prodan} of the coordinate and the projection operator. A particularly efficient method of calculating the Chern number has been proposed via the calculation of so-called coupling matrices\cite{yifu}.

Another method by which transitions between trivial and topological SPT states can be seen is through calculating the entanglement entropy\cite{lispt, osborne, xavier, canovi} (EE).  In a previous work we have shown that the EE of a clean system exhibits a cusp as a function of some model parameters at the point of a topological phase transition\cite{borchmann}. It should be made clear, however, that in SPTs the EE obeys the area law and it is this area-linear EE term which exhibits the cusp.  This should be contrasted with the case of system with intrinsic topology where a term referred to as 'topological entanglement entropy', $\gamma$, appears\cite{kitaev,Levin2}.  This term does not appear in SPTs.

In this letter we address the problem of disorder in a two dimensional topological superconductor (TSC).  Our TSC is a fully gapped, spin orbit coupled, superconductor in which time reversal symmetry is broken by a Zeeman field. It is therefore in class $D$.  We have studied this model previously in the clean limit and found topological phase transitions, which are evident from changes in the Chern number as well as the entanglement entropy cusps. 

We introduce disorder and search for topological phase transitions.  This is done in three ways (i)  by evaluating the Chern number in real space, (ii) by calculating the entanglement entropy and looking for a cusp when varying parameters and (iii) by calculating a disorder averaged self energy and using it to define an effective clean Hamiltonian for which the Chern number is easily found.  While topological phase transitions are found in all three ways, there are significant differences.  In particular, in the case of a $d$-wave superconductor with multiple gapped Fermi surfaces the real-space Chern number reveals a disorder induced topological phase. This phase appears in the real-space Chern number calculation as an intermediate phase where a single phase transition in the clean limit splits in to two transitions.  When using the self-consistent Born approximation to account for the renormalization of parameters one sees a hint of this intermediate phase, although its extent in parameter space is considerably smaller.  Surprisingly, the split into two phase transitions is not seen as a cusp in the entanglement entropy.

Our model is a two dimensional spin-orbit coupled topological superconductor on a square lattice with either $d+id$- or $s$-wave pairing symmetry.  We look at these two pairing symmetries due to their fundamentally different response to nonmagnetic impurities.
While in general, an $s$-wave\cite{anderson3} superconductor is robust against non-magnetic impurities, a $d$-wave superconductor is sensitive to this kind of scattering since its pairing amplitude depends on the momentum which is not conserved in the scattering process\cite{alloul,balatsky}.  This often leads to sub-gap states which, for a large number of impurities, can combine to form impurity bands. These sub-gap states can have a significant effect on the topology of the system\cite{kimme} when such a band crosses the Fermi surface and thus creates zero energy states. It has been shown that this can lead to gapless topological phases\cite{sau2} in disordered semiconducting nanowires.

\begin{figure*}[t]
\centering
\subfloat{\subfigimg[scale=.7]{a)}{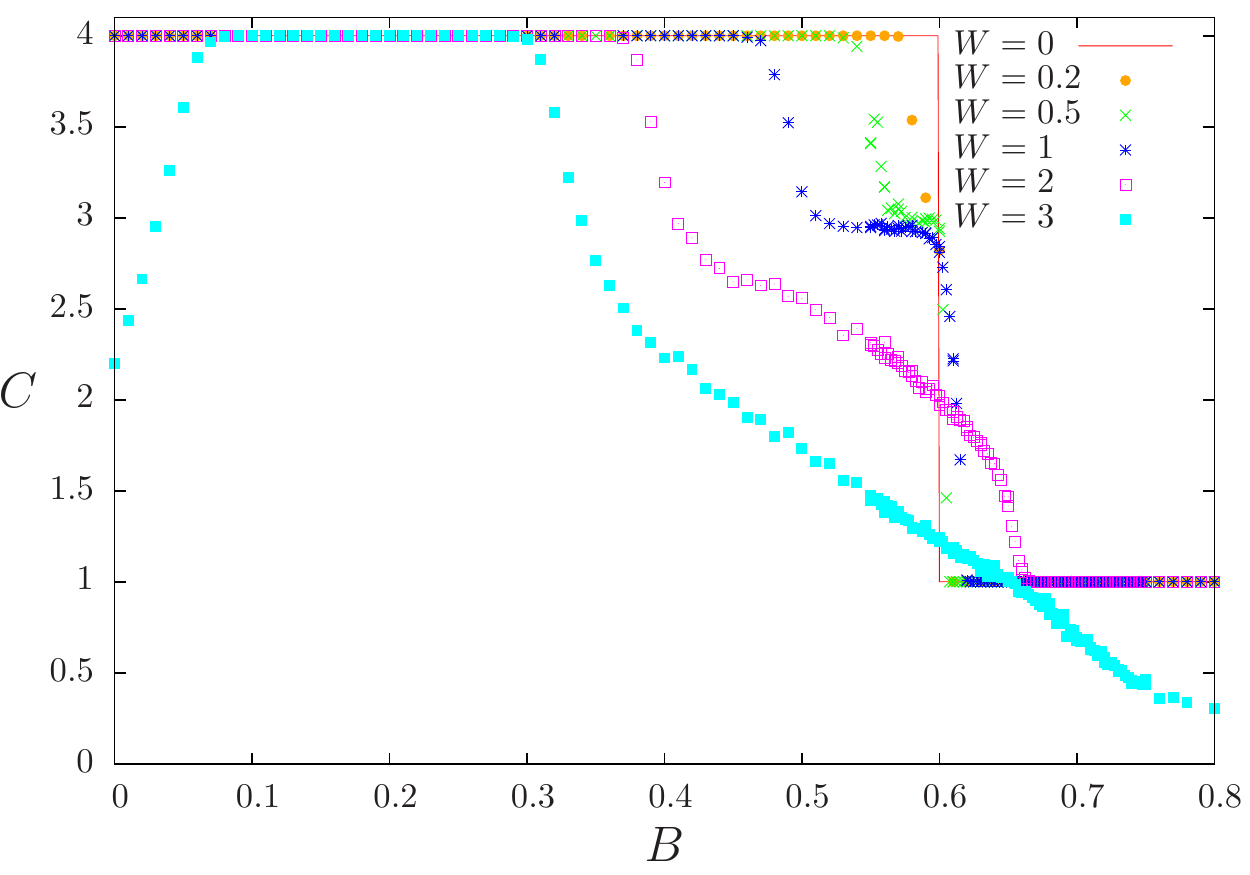}}
\subfloat{\subfigimg[scale=.7]{b)}{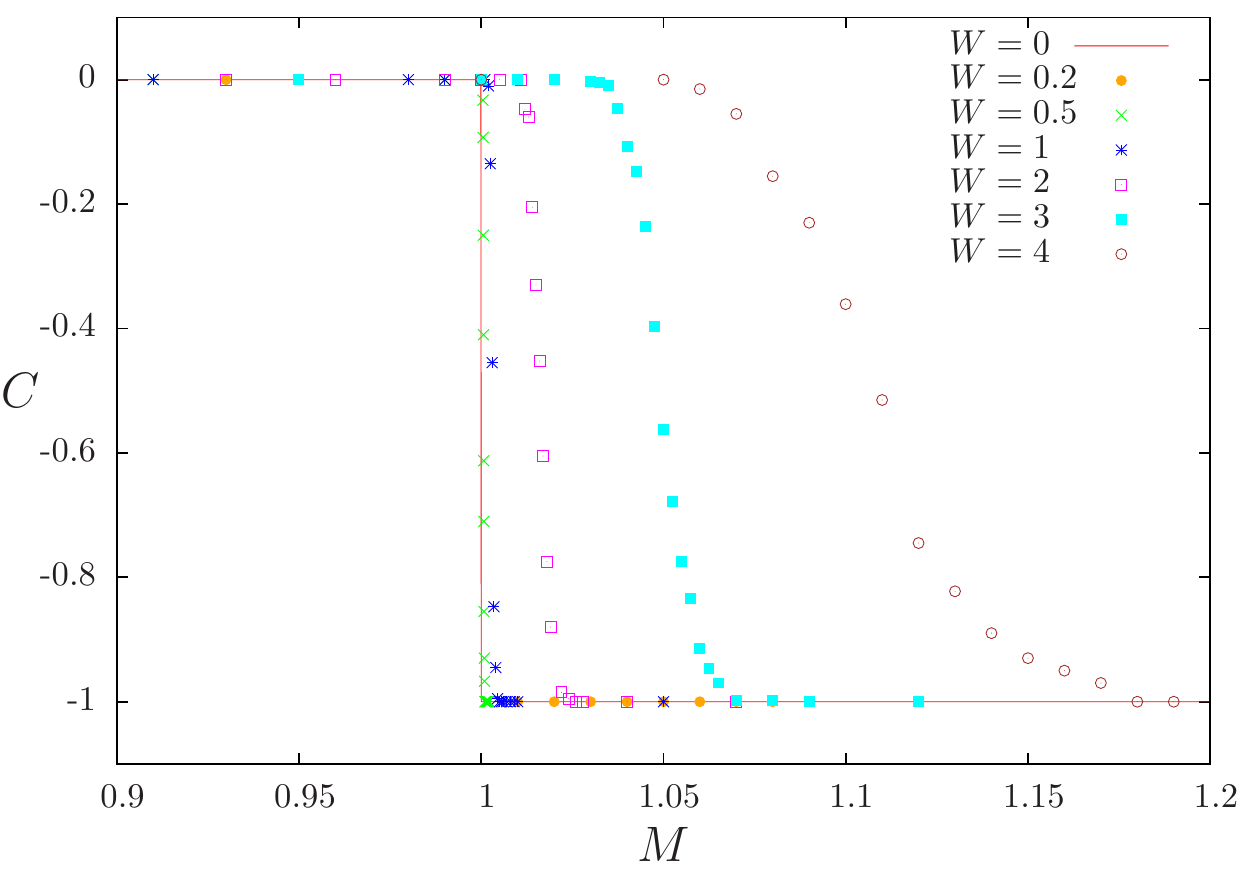}} \\
\subfloat{\subfigimg[scale=.7]{c)}{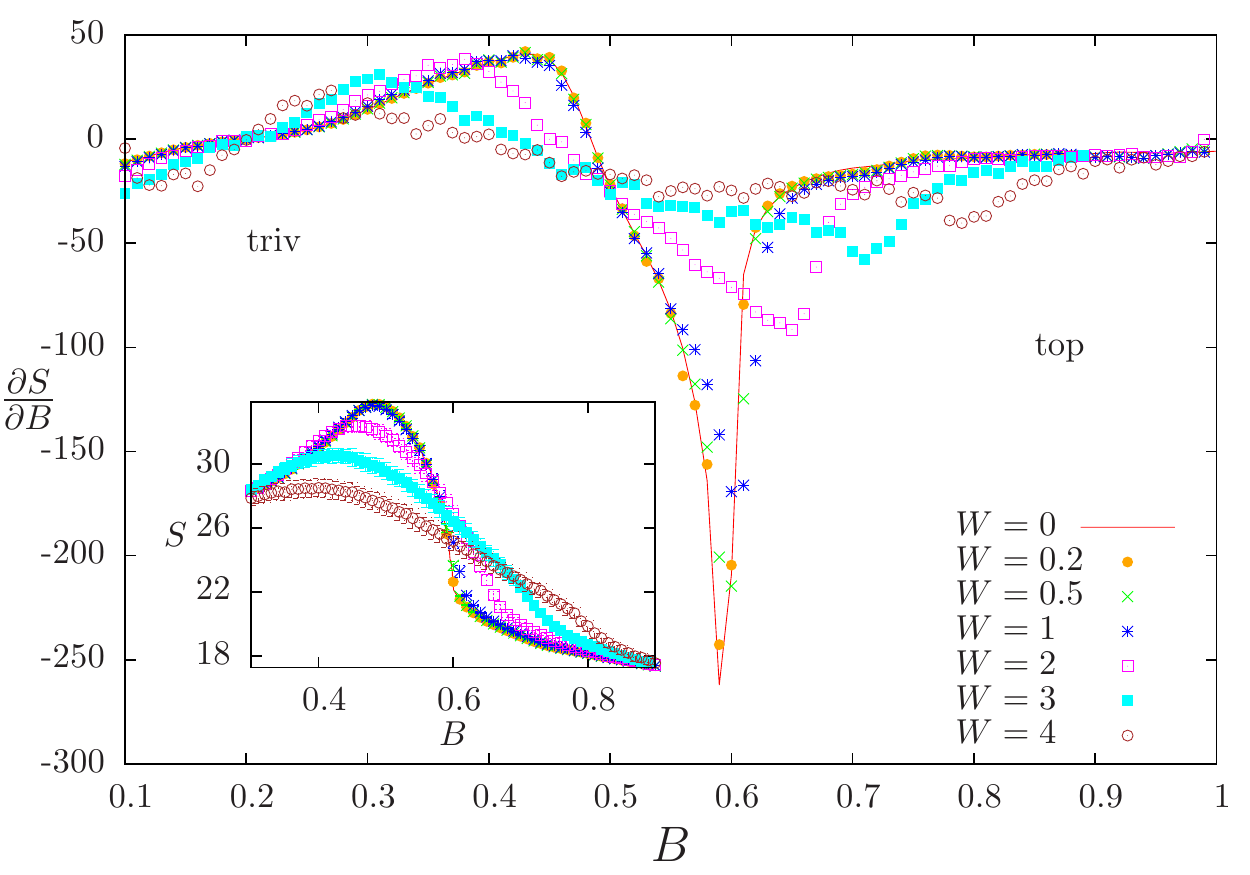}}
\subfloat{\subfigimg[scale=.7]{d)}{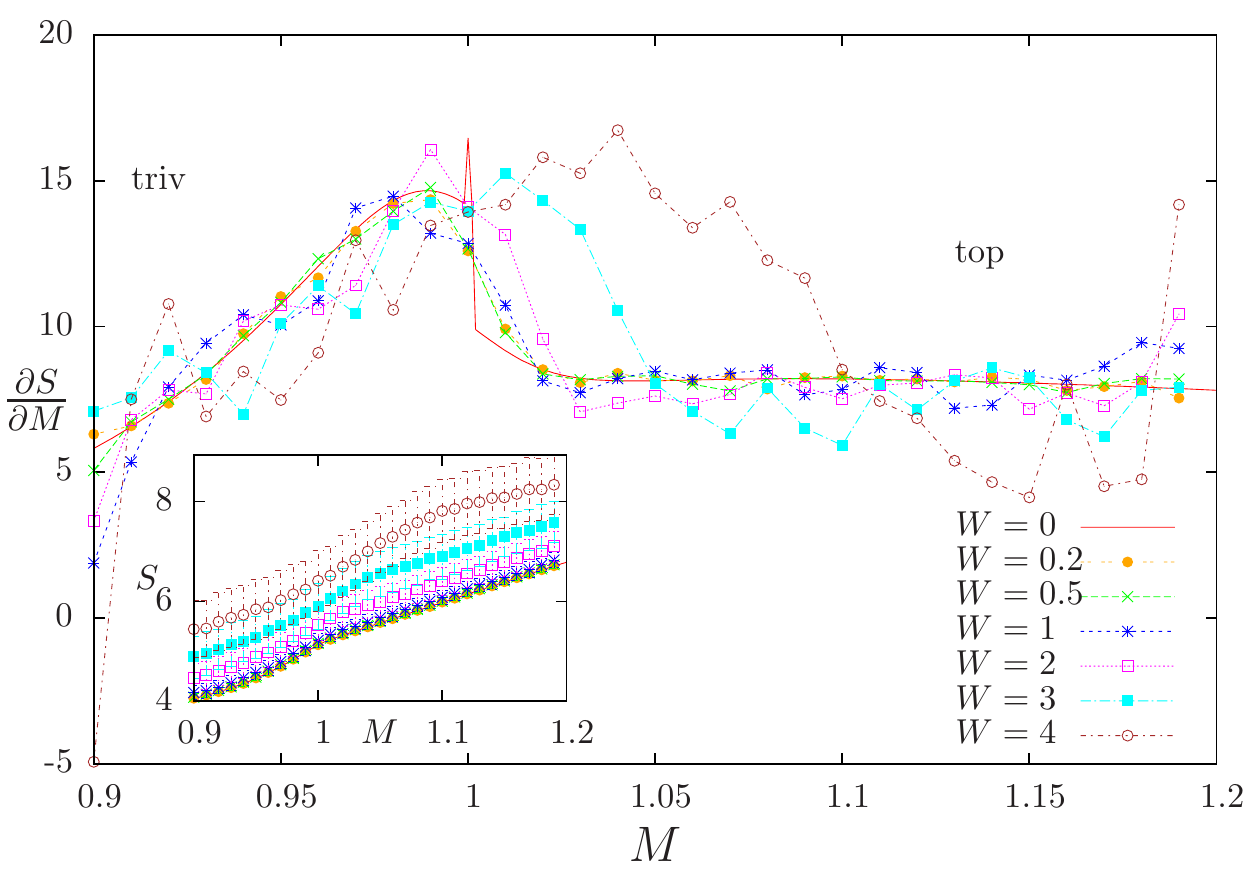}} 
\caption{(Color online) (a) Chern number $C$ for $d+id$-wave coupling $\Delta_1 = 0.8t,\Delta_2 = 0.4t,M=0.8t,\mu=0$ and $A=0.25t$, (b) Chern number $C$ for $s$-wave coupling $\Delta_s = 1t,B=0,\mu=-4t$ and $A=0.25t$. Derivative of the Entanglement entropy for (c) $d+id$-wave coupling , (d) $s$-wave coupling. The insets show the entanglement entropy.}\label{fig:entropy}
\end{figure*}
It should be noted that although the pairing term is of even angular momentum, when projected on to the spin-orbit coupled bands it acquires an additional phase winding.  This leads to effective $p$ or $f$-wave pairing in the bands\cite{Sau,Alicea}.  The question of whether the (clean) system is topological is therefore related to the number of spin-orbit coupled bands present.  If there is an odd number of bands this will lead to an odd Chern number. 

\section{Model} 
We use the Hamiltonian\cite{Farrell1} $H = H_K + H_{SO} + H_{SC} + H_D$,
where the kinetic energy part is given by nearest neighbour hopping,
\begin{align}
\begin{split}
H_K = -t \sum_{\langle i,j\rangle, \sigma} \left( c_{i\sigma}^\dagger c_{j\sigma} + c_{j\sigma}^\dagger c_{i\sigma} \right).
\end{split}
\end{align}
The spin-orbit part is given by
\begin{align}
\begin{split}
H_{SO} = \sum_\kv \psi_\kv^\dagger \left(  \sigma \cdot \mathbf{d}_\kv \right)\psi_\kv,
\end{split}
\end{align}
with $\psi_\kv = \left( c_{\kv\uparrow}, c_{\kv\downarrow} \right)^T$, $\sigma = (\sigma_x, \sigma_y, \sigma_z)$ and $\mathbf{d}_\kv = \left( A\sin{k_x}, A\sin{k_y}, 2B(\cos{k_x} + \cos{k_y} - 2) + M \right)$. Here, $A,B$ denote the strength of the Rashba and Dresselhaus spin-orbit coupling\cite{Sau}, respectively, and $M$ is the strength of the Zeeman term. The superconducting part is
\begin{align}
\begin{split}
H_{SC} = \sum_{\kv}\left( \Delta_{\kv}  c_{\kv,\uparrow} c_{-\kv,\downarrow} +\text{h.c.}\right),
\end{split}
\end{align}
where we look at two different pairing functions. For the fully gapped $d+id$-wave we have $\Delta_{\kv} = \Delta_1 (\cos(k_x)-\cos(k_y)) +i\Delta_2 \sin(k_x)\sin(k_y)$ and for the $s$-wave pairing $\Delta_\kv = \Delta_s$.

We include the effects of disorder by adding an on site, random potential term\cite{anderson4}
\begin{align}
\begin{split}\label{eq:D}
H_D = - \sum_i w_i c_i^\dagger c_i,
\end{split}
\end{align}
where $w_i \in [-\frac{W}{2},\frac{W}{2}]$ is a random number with a uniform distribution in the interval. $W$ is the overall disorder strength and a specific realization of the disorder is given by the set $\{ w_i \}$. When choosing and characterizing the size of the disorder strength, we are guided by the typical energy scales of the system, the gap and the bandwidth. Comparing with the gap, c.f. Fig. 1c in Ref. \onlinecite{borchmann}, one can see that $W=1$ is larger than the gap. $W=3$ is of the order of the bandwidth. We compute disorder averaged quantities, namely, the Chern number and EE, by calculating the quantity for a specific realization of the disorder and then averaging over a large number ($\ge 400$) of realizations $\{ w_i \}$.  The number of disorder realizations is taken such that the average quantities and standard deviations have saturated and do not change upon including more disorder realizations.  We find that for low disorder strength 400 realizations are sufficient while for higher disorder we need larger samples. 


\section{Real Space Chern number} 
In order to analyze the behaviour of the system in the presence of disorder, one can calculate the topological invariant of the ground state by using a real space formula\cite{yifu, prodan}. This formula is derived by writing the wavefunctions on the torus and constructing their Fourier components with twisted boundary conditions in both directions.  The Chern number can then be evaluated as the response to the twists. By using the twisted boundary conditions, the ground states induces the structure of a $U(1)$-fibre bundle over the torus of phase twists, whose Chern number gives the topological invariant.

In order to make a connection with our previous work\cite{borchmann}, we look at the phase transition in the $d$-wave system which, in the clean limit, takes place at the value $B_c = 0.6t$ (and the other parameters are set to $M=0.8t,\mu=0$ and $A=0.25t$).  In the clean system, for $B<B_c$ the superconductor is trivial\footnote{We call this phase trivial due to the fact that the even number of edge modes will hybridize due to the broken time reversal symmetry.} with $C=4$ and for $B>B_c$ it is a topological superconductor with a Chern number of  $C=1$. 

In Fig.~\ref{fig:entropy}a) we show the result of the Chern number calculation in real space for the clean system and for various disorder strengths in the $d$-wave system.
Looking at the graph, the first striking feature is that compared with the clean system, the disordered system has an additional phase.
While in the clean limit one finds a single, sharp transition between a trivial $C=4$ phase on the left to a $C=1$ phase on the right, for disorder strength of $W=0.5t$ to $1t$ the transition splits to two and an intermediate phase with $C=3$ appears.  The transition from $C=4$ to $C=3$ appears before (for lower B) the clean limit transition and does not cause a cusp in the entanglement entropy.  The second transition from $C=3$ to $C=1$ occurs after the clean limit transition and shows as a cusp in the EE.  At the $C=3$ plateau a large majority of the disordered systems, ranging from 65\% up to 95 \%, have a Chern number of $3$, while a small fraction have $C=4$ or $C=2$ moving the average slightly away from $3$.  At the two other phases, with $C=4,1$ all of the systems in the average have exactly the same Chern number. 

For $W=3t$ the disorder averaged Chern number does not saturate to 1 anymore. This is caused by the fact that the system becomes gapless\cite{sau2} and the Chern number is no longer well defined. Specifically, this behaviour implies the vanishing of the mobility gap as localized states do not influence the Chern number.  Consequently, the real space Chern number is not a good indicator of the topology of the system in this regime. Furthermore, for low $B$ the Chern number starts deviating from its clean value due to the fact that it is sensitive to another phase transition taking place at $B=-0.4t$.

One can speculate on the origin of the new disorder-induced topological phase. First, a Chern number of $4$ is an indication that multiple Fermi surfaces contribute to the topological invariant. Therefore it is possible that the change in Chern number does not occur simultaneously in all Fermi surfaces.  Moreover, one can imagine that localized states may reduce the life time of the bands and change the overall topological nature. Indeed, a disordered induced topological phase is not seen in the $s$-wave superconductor where potential disorder is not expected to cause localization. We should also note that a similar effect of localization was encountered in the case of symmetry class DIII\cite{kimme}.   

The disorder averaged Chern number in the $s$-wave system is shown in Fig.~\ref{fig:entropy}b. In general, for this pairing symmetry, the system only exhibits phase transitions with $\Delta C = \pm 1$.  We choose to focus on one of these transitions, which is controlled by the model parameter $M$. Note that due to the momentum independence of the $s$-wave order parameter, no pair breaking is induced by the disorder and no sub-gap states appear.  We therefore expect the effect of disorder in this superconductor to be different from that of the $d+id$ case. 

\section{Disorder Averaged Entanglement Entropy} 
Several authors have studied the entanglement properties of disordered systems\cite{mondragon,andrade,pouranvari,mondragon2,vijay}.  In particular, the relation between the level spacing in the entanglement spectrum and the topology was explored in Refs.~[\onlinecite{prodan, gilbert}]. In the current work we focus on the entanglement entropy of disordered SPTs and investigate whether a topological phase transition is seen as a kink in the EE as was seen in the clean limit\cite{borchmann}.  We follow the above kink as the strength of the disorder is increased.

The disorder averaged entanglement entropy can be defined as the disorder averaged von Neumann entropy of the reduced density matrix, $\overline{S_A} = \overline{\text{Tr}\left( \rho_A\ln{\rho_A} \right)}$, where $A$ is a partition of the original system. For our calculations we define $A$ as a 12x12 square in a 40x40 lattice with periodic boundary conditions, where the remaining degrees of freedom of the original system were traced out. We calculate the reduced density matrix via the two-point correlation function\cite{peschel1}. The size of the system is limited by the fact that these calculations are done in real space as well as the need for statistical averaging. 

In Fig.~\ref{fig:entropy}c we show the entanglement entropy of the $d$-wave model (inset) and the derivative of the entropy with respect to the parameter $B$.  While the EE exhibits a cusp at the phase transition, it is easier to recognize the transition in the derivative.  The figure shows that the clean system's sharp transition at $B_c$ is shifting to higher values of $B$ when the disorder is increased and becomes less sharp at the same time.  For strong disorder the transition is completely washed out. One can see that the derivative of the EE displays only a single kink, coinciding with the position of the phase transition from $C=3$ to $C=1$, while any signature of the first phase transition is completely absent.

In Fig.~\ref{fig:entropy}d the EE for the $s$-wave system is presented as a function of the parameter $M$.  We note that in this system the transition is not as pronounced as in the $d$-wave case even in the clean limit.  When following the transition we see that once the disorder is applied the transition moves to higher values of $M$ and its position coincides well with the one obtained via the real space Chern number. It also becomes less sharp and washes out completely for strong disorder.


\section{Disorder Averaged Self Energy} 
Another approach that is often used to deal with disordered systems is using the Gaussian disorder properties to define an averaged Green's function and restore the translation symmetry. In other words, the disorder induces a self energy which renormalizes the model parameters.  In the case of an Anderson topological insulator it was shown that a Gaussian disorder leads to a change in the Zeeman field parameter ($M$ in our model) which in turn leads to a change in topology.\cite{groth}  We therefore apply the same method here. 

To this end, we use the variance of the random potential above and write, $\overline{V(q)V(q')} = \frac{W^2}{12\mathcal{V}}\delta_{q+q',0}$, where $\mathcal{V} = L^2$ is the volume of our ($L\times L$ lattice) system. With this, the self energy in the self-consistent Born approximation (SCBA) reads
\begin{align}
\begin{split}
\Sigma(\omega) &= \frac{W^2}{12\mathcal{V}} \times \\
&\sum_q \left( \sigma_0 \otimes \tau_z \right) \cdot \left(\omega - H(q) - \Sigma(\omega) + i\eta\right)^{-1} \cdot \left( \sigma_0 \otimes \tau_z \right)
\end{split}
\end{align}
where $\tau_i$ are Pauli matrices acting on the particle and hole degree of freedom and $\sigma_i$ denote the spin. The self consistent summation includes all non-crossing diagrams.
Focusing on the static limit we can think of $\Sigma = {\rm Re}\left(\Sigma(\omega=0)\right)$ as renormalizing the parameters of the Hamiltonian.  Consequently we can define the effective Hamiltonian\cite{wang, Witczak-Krempa} $H_\Sigma = H + \Sigma$ in which one can easily calculate the relevant Chern number in momentum space.\cite{ghosh, Farrell2}  Due to the parameter renormalization the topological phase transition moves in parameter space with respect to the clean system.
The fact that $\Sigma$ is independent of momentum, limits the possible quantities that can be renormalized to $M,\mu$ as well as $\Delta_s$ in the $s$-wave pairing case. Thus, most generally this can be written as\cite{alloul,balatsky}
\begin{align}
\begin{split}
\Sigma = -\mu_R \left( \sigma_0 \otimes \tau_z \right) + M_R \left( \sigma_z \otimes \tau_z \right) - i\Delta_{sR} \left( \sigma_y \otimes \tau_x \right),
\end{split}
\end{align}

\begin{figure*}
\centering
\subfloat{\subfigimg[scale=.7]{a)}{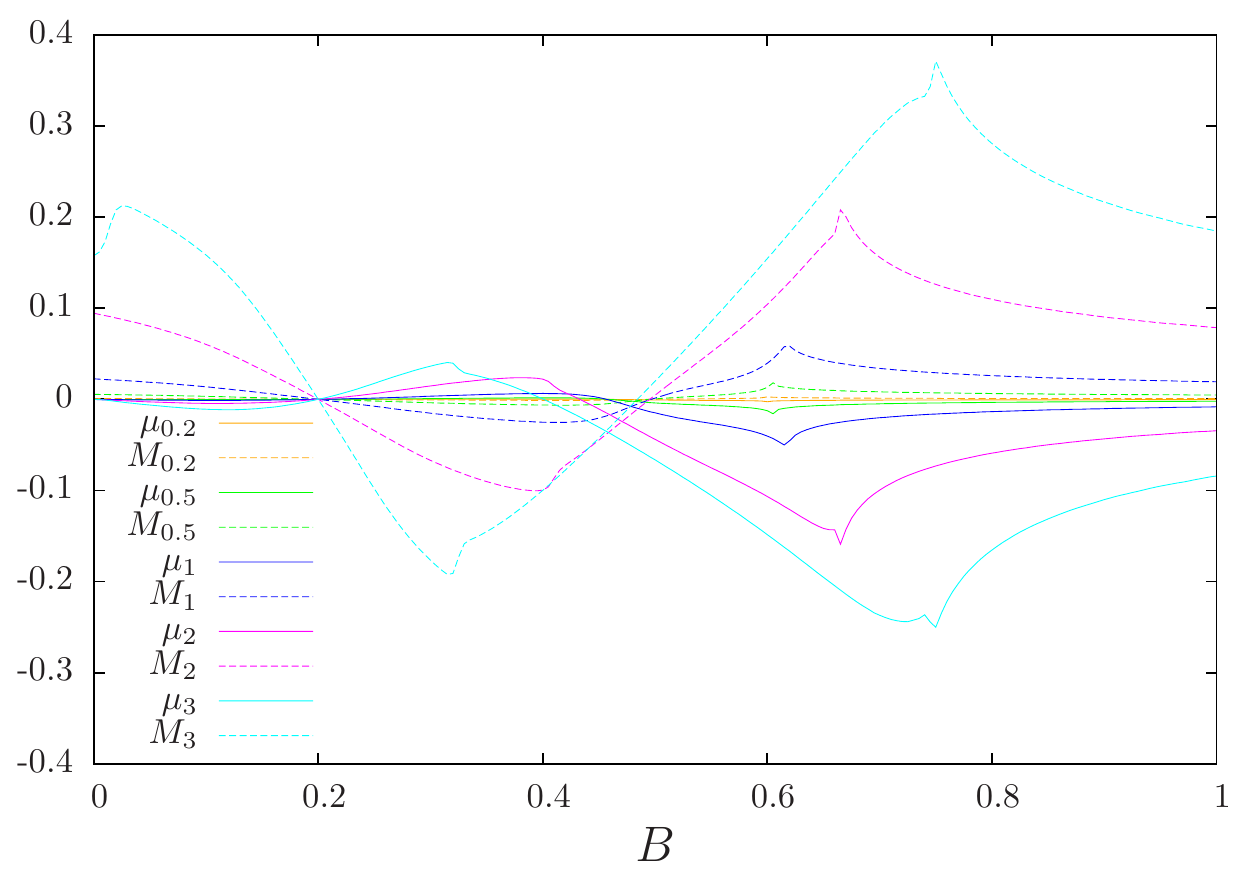}}
\subfloat{\subfigimg[scale=.7]{b)}{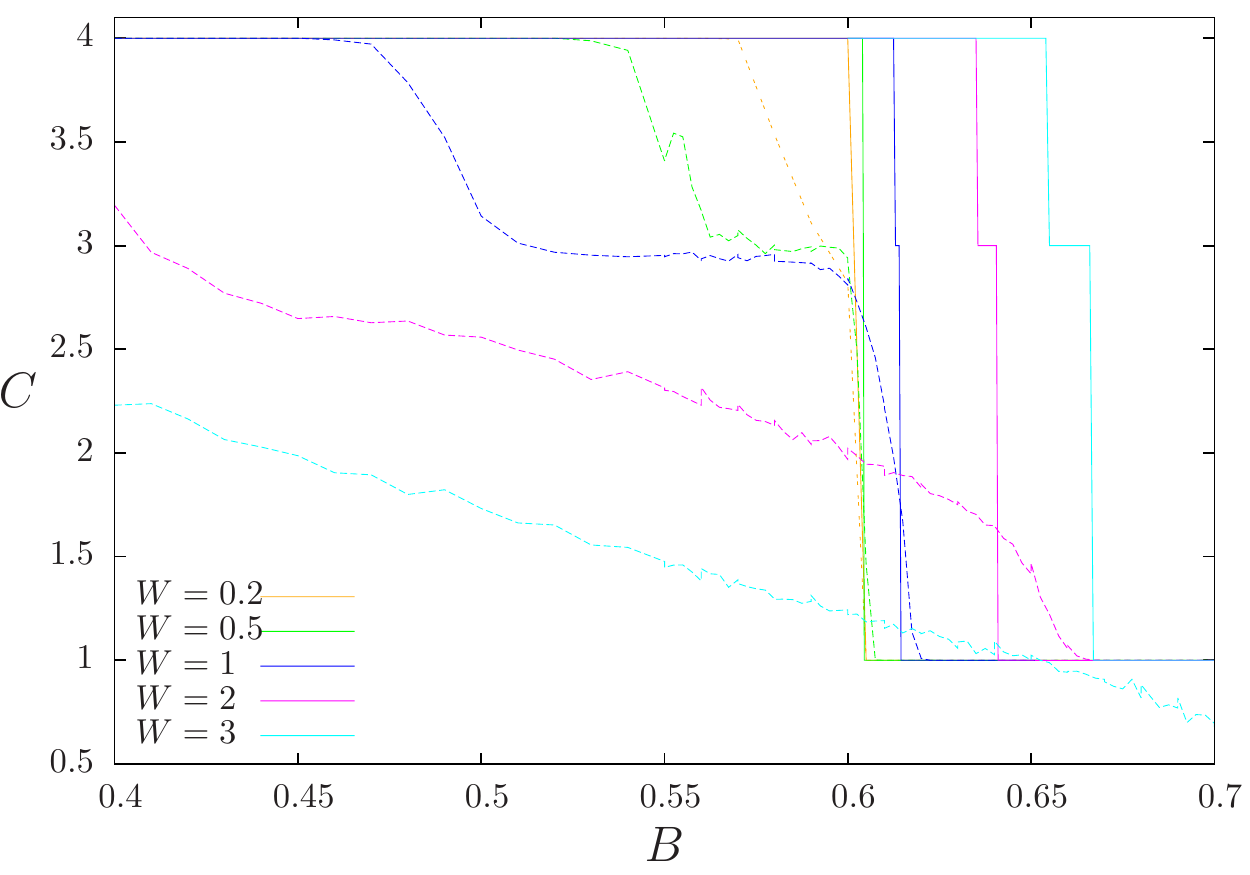}}
\caption{(Color online) (a) Renormalization of the chemical potential (solid line) and Zeeman coupling (dashed line) for varying disorder strengths for $d+id$-wave and (b) Chern number $C_{ren}$ (solid line) calculated from the renormalized parameters and through a real space formula $C_{RS}$ (dashed line).}\label{fig:renorm}
\end{figure*}
where the renormalized parameters are $M_R = M + M_\Sigma$, $\mu_R = \mu+\mu_\Sigma$ and $\Delta_{sR} = \Delta_s+\Delta_\Sigma$. 
Looking at the results in Fig.~\ref{fig:renorm} we see that there is good agreement between this method and the real-space Chern  number calculation as well as the EE cusp with respect to the transition between the $C=3$ and the $C=1$ phases.  On the other hand, the first transition, from $C=4$ to $C=3$ which appears in the real-space Chern number, appears in the self energy at a higher $B$ value and is completely absent from the EE.  Overall, the intermediate $C=3$ phase appears in the self-energy calculation but its range is smaller by about an order of magnitude than its range in the real-space Chern number calculation.  We note that the chemical potential renormalization is the most important one when it comes to creating the $C=3$ phase.  We speculate that the SCBA, which neglects cross diagrams, might not be sufficient when estimating the $C=3$ phase range.

%

\section{Conclusion} 
In this work we have presented evidence for a disorder induced topological phase for certain ranges of disorder strength. 
The calculation of the real space Chern number as well a disorder averaged self-energy predict the appearance of a new, $C=3$-phase between the $C=4$ and the $C=1$ phase which exist in the clean system. However, the range of parameters over which the disorder induced phase occurs is much smaller in the self-energy method compared with the real-space Chern number.  This is perhaps a result of the self-consistent Born approximation which neglects cross diagrams.
In addition, we find that the disorder averaged entanglement entropy is a useful indicator in some topological phase transitions but not others. In particular, in the $d$-wave case, it has a cusp in the transition between the new, disorder induced topological phase and the $C=1$ phase but does not have a cusp at the transition between the $C=4$ and $C=3$ phase.

\section{Acknowledgements} 
The authors are grateful for useful discussions with M.~Franz, L.~Fu and B.~Seradjeh. Financial support for this work was provided by the NSERC and FQRNT (TPB, JB) and the Vanier Canada Graduate Scholarship (AF). The majority of the numerical calculations were performed using CLUMEQ/McGill HPC supercomputing resources.  

\bibliographystyle{apsrev}
\bibliography{topoSC}

\end{document}